\documentclass[prl,twocolumn,showpacs,preprintnumbers,amsmath,amssymb,superscriptaddress]{revtex4}
\usepackage{graphicx}
\usepackage{dcolumn}
\usepackage{bm}
\usepackage{psfrag}

\begin{document}
\title{Theory of antibound states in partially filled narrow band systems} 
\author{G. Seibold}
\affiliation{Institut f\"ur Physik, BTU Cottbus, PBox 101344,
03013 Cottbus, Germany}
\author{F. Becca}
\affiliation{INFM-Democritos, National Simulation Center and International
School for Advanced Studies (SISSA), I-34014 Trieste, Italy}
\author{J. Lorenzana}
\affiliation{SMC-INFM, ISC-CNR, Dipartimento di Fisica,
Universit\`a di Roma ``La Sapienza'', P. Aldo Moro 2, I-00185 Roma, Italy}
\date{\today}

\begin{abstract}
We present a theory of the dynamical two-particle response function in the 
Hubbard model based on the time-dependent Gutzwiller approximation. The results 
are in excellent agreement with exact diagonalization on small clusters and
give reliable results even for high densities, where the usual ladder 
approximation breaks down. We apply the theory to the computation of
antibound states relevant for Auger spectroscopy and cold atom
physics. A special bonus of the theory is its computational simplicity. 
\end{abstract}


\maketitle
Much of our understanding of strongly correlated electronic systems comes from
dynamical responses like the one-particle spectral function  
which, under certain approximations, is probed by
photoemission and inverse photoemission experiments. 
Less explored is the two-particle spectral function in which one studies 
how the system responds to the addition or removal of two particles. 
In the case of two holes in an otherwise filled band, the response was computed
exactly by Cini\cite{cin76,cin78} and Sawatzky\cite{saw77} (CS) in
connection with Auger spectroscopy. They showed that for strong enough
on-site repulsion the spectral function gets dominated by antibound 
states in which the two holes propagate together paying a large Coulomb cost.  

Despite the interest of the problem, the dynamical two-particle response and 
the formation of antibound states in {\em partially-filled} correlated systems 
are not well understood. Cini and collaborators\cite{cin86,ver01} have compared
approximations for the spectral function developed by several groups with exact
diagonalization on finite clusters. They observed that any attempt to improve 
the single-fermion propagators with self-energy corrections or making them self
consistent leads to worse results due to the lack of vertex corrections which, 
if included, would tend to ``undress'' the Green's functions. Thus for small 
filling, the best approximation corresponds to a trivial generalization of the 
original theory, namely summing a ladder series with bare Green's functions.
For moderate filling and for large interactions, this bare 
ladder approximation (BLA) breaks down and no reliable theory is available.  

Several effects are expected to be relevant in the case of a partially-filled 
band. First, strong correlation produces band narrowing, which should help 
to split-off antibound states from the two-particle continuum. Second, the 
spectral weight of the antibound state should depend on doping, since the 
probability to find an empty site where to create an antibound pair depends
on the filling. Third, the other holes present in the
system are expected to screen the effective interaction among the
added holes, which may lead to a renormalization of the position of the
antibound state with respect to the continuum. Last, the chemical
potential has a jump as a function of doping across the Mott insulating phase 
of narrow-band systems, which should show up in the position of the 
two-particle continuum with respect to the antibound state. 

\begin{figure}[tb]
\includegraphics[width=8.5cm,clip=true]{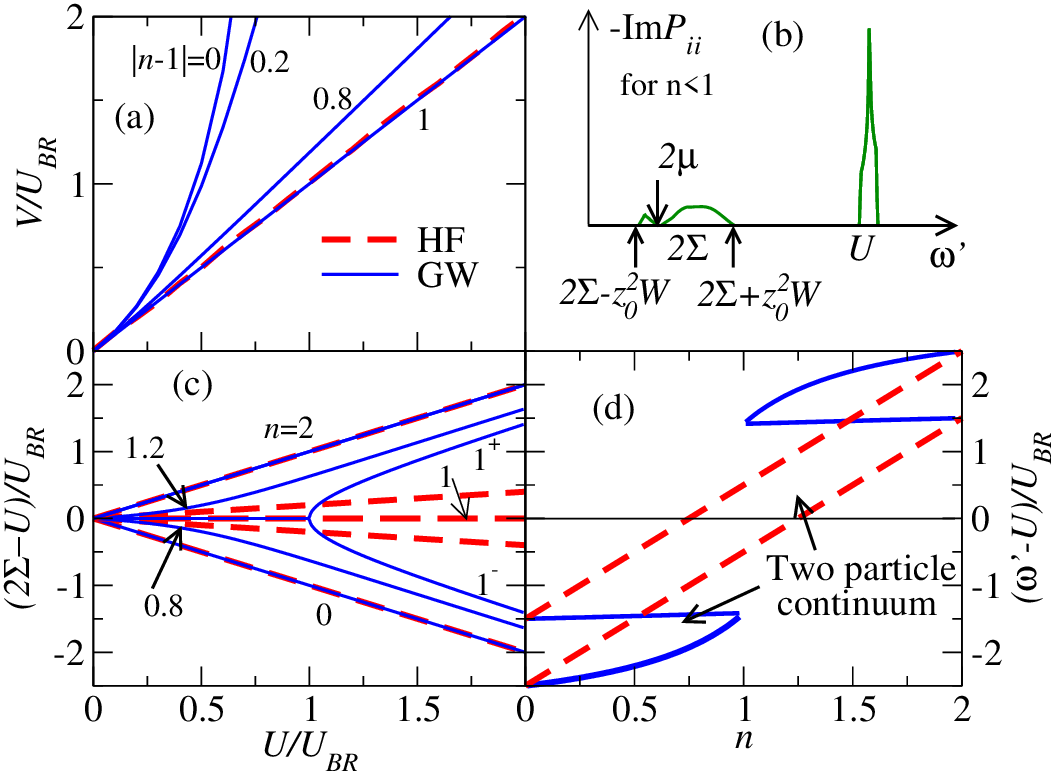}
\caption{
Panels (a), (c), and (d) are obtained within 
a model with a flat density of states and a bare bandwidth $W$
($U_{BR}\equiv 2 W$) in HF (dashed line) and GA (full lines).
(a) Effective particle-particle interaction $V$  
for different dopings $n-1$. Negative and positive dopings coincide. 
(b) Sketch of the principal energy scales of ${\rm Im}P_{ii}$. 
(c) Energy distance between the center of the two-particle scattering states 
($2\Sigma$) and the doublon energy $U$ for different fillings.  
The $1^+,1^-$ fillings are infinitesimal deviations from half filling and 
coincide in HF and in GA for $U<U_{BR}$. The same plot represents the 
effective interaction rescaled by the doping $V(n-1)$ 
[see Eq.~\eqref{eq:veff}]. 
(d) Boundaries of the two-particle continuum 
$\omega'=2\Sigma\pm z^2_0 W$ for $U=2U_{BR}$ as a function of band filling.
}
\label{figueff}
\end{figure}

\begin{figure}[tb]
\includegraphics[width=8cm,clip=true]{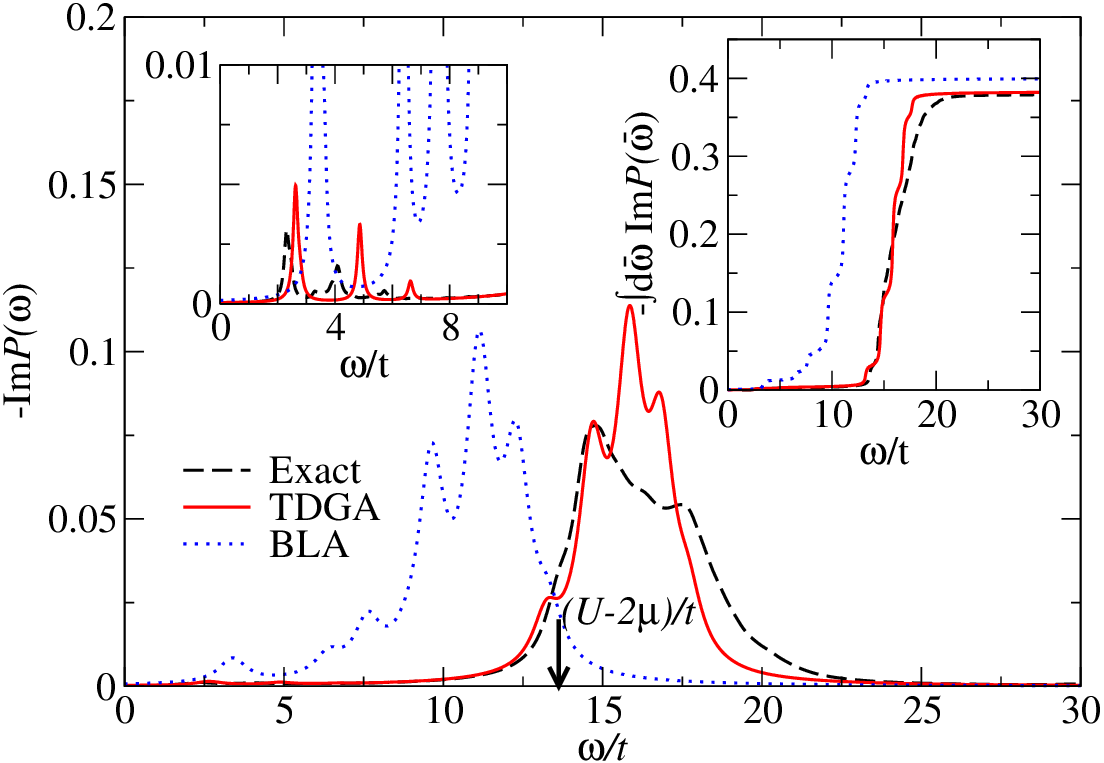}
\caption{
Imaginary part of the pair correlation function for the Hubbard model with 
$U/t=15$ and $10$ particles on a $4\times 4$ square lattice obtained by exact 
diagonalization, TDGA and BLA. The origin of energy is at $2\mu$. 
The upper-left inset enlarges the region of low-energy band excitations and 
the upper-left inset shows the frequency evolution of the integrated spectra. 
The broadening of the delta-peaks is $0.5t$ in the main panel and $0.1t$ in 
the insets. The arrow indicates the exact $U-2\mu$ value.}
\label{fig:u15}

\vspace*{-0.4cm}

\end{figure}

In this work, we present a theory of antibound states for the Hubbard model, 
which incorporates these effects. It is based on the computation of pairing 
fluctuations within the time-dependent Gutzwiller approximation 
(TDGA)\cite{sei01}. Our approach reproduces the effects discussed above, while
keeping the simplicity of CS theory. Interestingly, we find 
that the effect of a finite density is to antiscreen the Hubbard $U$ 
interaction, i.e., the effective interaction is larger than the bare one and 
becomes singular as the Mott phase is approached 
[c.f. Fig.~\ref{figueff}(a)]. The comparison of our results with 
exact diagonalizations shows that TDGA is reliable even at high densities 
where the BLA breaks down (c.f. Fig.~\ref{fig:u15}).

Our starting point is the Hubbard Hamiltonian:
\begin{equation}\label{HM}
\hat H=\sum_{ij\sigma}(t_{ij}-\mu \delta_{ij})
c_{i\sigma}^{\dagger}c_{j\sigma} + U\sum_{i}
n_{i\uparrow}n_{i\downarrow}
\end{equation}
where $c^\dagger_{i\sigma}$ creates a fermion with spin $\sigma$ at site $i$, 
$n_{i\sigma}=c_{i\sigma}^{\dagger}c_{i\sigma}$, $U$ is the on-site 
repulsion, $t_{ij}$ denotes the hopping amplitude, and $\mu$ is the chemical 
potential. 

We are interested in the following two-particle response
\begin{equation}\label{eq:cross}
P_{ij}(\omega)= \frac{1}{\imath} \int_{-\infty}^{\infty}dt 
\mbox{e}^{\imath\omega t} \langle {\cal T} c_{i\uparrow}(t) c_{i\downarrow}(t)
c_{j\downarrow}^\dagger c_{j\uparrow}^\dagger \rangle,
\end{equation}
for $\omega>0$ ($\omega<0$) the imaginary part of Eq.~\eqref{eq:cross} gives 
the two-particle addition (removal) spectra. For the 
Auger application one should consider other 
effects which have been extensively discussed in the 
literature\cite{ver01} and will not be treated here (e.g.,
finite life time of the core hole\cite{gun80} and the interaction of
the core hole with the valence electrons\cite{pot93}). 

Our approach is based on the Gutzwiller wave function\cite{gut65,vol84}:
$|\Phi\rangle =P_g | \phi \rangle $ where $P_g$  partially projects out doubly 
occupied sites from $|\phi \rangle $, which we assume to be a Bogoliubov 
vacuum. We define the single-particle density matrix 
$\rho_{i\sigma j\sigma'}\equiv \langle  \phi|c_{j\sigma'}^{\dagger}
c_{i\sigma} | \phi \rangle$ and pair matrix 
$\kappa_{i\sigma, j\sigma'}\equiv\langle \phi|c_{j\sigma'}
 c_{i\sigma} | \phi \rangle$, which satisfy the following 
constraints\cite{bla86}
\begin{equation}\label{eq:cons}
\rho^2-\rho=\kappa\kappa^*,\;\;\; \;\;\;\;\;\;\;  [\rho,\kappa]=0. 
\end{equation}

The first step is to construct the charge rotationally invariant
energy functional $E\equiv \langle\Phi|H|\Phi\rangle$
in the Gutzwiller approximation (GA). 
This is more easily done by rotating at each site the fermion 
annihilation and creation operators to a basis where the anomalous 
expectation values vanish\cite{sof92}. Then, one derives the GA with one of 
the known techniques\cite{kot86,geb90} and rotates back to the original 
 operators. Restricting to a paramagnetic state one finds
\begin{equation}\label{eq:ej}
E[\rho,\kappa,D]=\sum_{ij\sigma} t_{ij}
z_i z_j \rho_{i\sigma,j\sigma} 
+ U\sum_{i} D_i,
\end{equation}
with the hopping renormalization factors
\begin{equation}\label{qfac}
z_{i}=\frac{\sqrt{\frac12-D_i+J_{iz}}
\left( \sqrt{D_i-J_{iz}-J_i}+ \sqrt{D_i-J_{iz}+J_i}\right)}
{\sqrt{\frac14-J_i^2}}.\nonumber
\end{equation}
Here we defined $J_{ix}=(\kappa_{i\uparrow,i\downarrow}+
\kappa_{i\uparrow,i\downarrow}^*)/2$, 
$J_{iy}= \imath (\kappa_{i\uparrow,i\downarrow}-
\kappa_{i\uparrow,i\downarrow}^*)/2$, 
$J_{iz}=(\rho_{i\uparrow,i\uparrow}+\rho_{i\downarrow,i\downarrow}-1)/2$,
$J_i\equiv |{\bf J}_i|$, and the double occupancy 
$D_i = \langle\Phi|n_{i\uparrow}n_{i\downarrow}|\Phi\rangle$.
The ground state is found by minimizing Eq.~\eqref{eq:ej} with the
constraints~\eqref{eq:cons}, leading to the static $\rho^0$, 
$\kappa^0$, ${\bf J}^0$ and $D^0$. We will consider a paramagnetic normal
metal thus $\kappa^0=J_x^0=J_y^0=0$.

To compute the response function we add a weak time-dependent pairing field
$F(t)=\sum_{i} (f_{i} e^{-\imath \omega t} c_{i\downarrow} c_{i\uparrow} 
+h.c.)$ to Eq.~\eqref{HM}. This produces small time-dependent deviations   
$\delta \rho(t)=\rho(t)-\rho^0$. In addition, since $F$ does not conserve the
particle number, it induces pairing correlations $\kappa$, which we 
compute in linear response.

Previously\cite{vol84,sei01,sei03}, the energy 
was expanded to second order in terms of particle-hole fluctuations, leading 
to effective matrix elements for charge and spin excitations. 
For a normal paramagnet neither those channels nor 
$\delta D$ fluctuations mix with the particle-particle channel which
simplifies the formalism. The remaining part follows text book computations 
in nuclear physics\cite{bla86}. 

Expanding the energy up to second order in $\delta \rho$ and $\kappa$ one finds:
\begin{equation}\label{eq:ener}
\delta E= \sum_{{\bf k}\sigma}(\varepsilon_{\bf k}-\mu) \delta\rho_
{{\bf k}\sigma,{\bf k}\sigma}+ V \sum_i (J_{ix}^2+J_{iy}^2).
\end{equation}
Here $\varepsilon_{\bf k}\equiv z_0^2 e_{\bf k}+\Sigma_G$ denotes the 
GA dispersion relation ($e_{\bf k}$ is the bare one),  
$\Sigma_G$ coincides with the Lagrange parameter of the slave boson
method\cite{kot86} and is given by $\Sigma_G=z_0 z_0' \bar e$ with 
$\bar e \equiv\sum_{i\sigma} t_{ij}\rho_{i\sigma j\sigma}^0$,
$z_0$ is the hopping renormalization factor at the saddle point and $z_0'$ is 
its density derivative. Our notation emphasizes the fact that $\Sigma_G$ can 
be interpreted as a local GA self-energy. 
Finally, the effective on-site particle-particle interaction is
\begin{equation}\label{eq:veff}
V = \frac{U-2 \Sigma_G}{1-n},
\end{equation}
where $n$ denotes the particle concentration.  At half filling ($n=1$), both 
the numerator and the denominator tend to zero and one 
finds $V=U\left( 1-U/2U_{BR} \right) (1+U/U_{BR})/(1-U/U_{BR})$, which 
coincides with the particle-hole case\cite{vol84,sei01,sei03}. 
Here $U_{BR}=8 \bar e$ is the critical interaction for the
Brinkman-Rice metal insulator transition\cite{bri70,vol84}. 

The response function can be readily derived from the equations of motion of 
the pair matrix in a normal system after using the constraint~\eqref{eq:cons} 
to express the first term in Eq.~\eqref{eq:ener} as a quadratic contribution 
in $\kappa$\cite{bla86}. The momentum dependent pair-correlation function is 
given by the usual ladder expression but with the effective interaction of
Eq.~\eqref{eq:veff}:
\begin{equation}\label{eq:prpa}
P({\bf q},\omega) = \frac{P^0({\bf q},\omega)}{1-V P^0({\bf q},\omega)},
\end{equation}
where $P^0({\bf q},\omega)$ is the non-interacting two-quasiparticle
correlation function 
\begin{equation}\label{chi0}
P^0({\bf q},\omega)= \frac{1}{N_s} \sum_{\bf k} \frac{1-f(\varepsilon_{\bf k}) 
- f(\varepsilon_{{\bf k+q}})}{\omega - \varepsilon_{\bf k} -
\varepsilon_{{\bf k+q}}+2\mu+ \imath \eta_{{\bf k},{\bf k+q}}}
\end{equation}
evaluated with the GA dispersion relation $\varepsilon_{\bf k}$. $N_s$ is the 
number of sites, $f(\varepsilon_{\bf k})$ is the Fermi distribution function, 
and $\eta_{{\bf k},{\bf k'}}\equiv 0^+ {\rm sign}(\varepsilon_{\bf k}+\varepsilon_{{\bf k}'}-2 \mu)$. 

Eqs.~\eqref{eq:veff},\eqref{eq:prpa}, and~\eqref{chi0} constitute our main 
result. Our approach leads to the same formal ladder structure as in the
CS theory\cite{cin76,cin77,saw77} but with the HF self-energy 
($\Sigma_{HF}\equiv U n/2$, $z_0=1$) replaced by the GA one and the Hubbard 
repulsion $U$ replaced by an effective interaction $V$. Notice that 
 the ``new'' Eq.~\eqref{eq:veff} is valid in the BLA provided
 one replaces $\Sigma_{G}\rightarrow  \Sigma_{HF}$ leading to $V=U$.

\begin{figure}[tb]\includegraphics[width=8cm,clip=true]{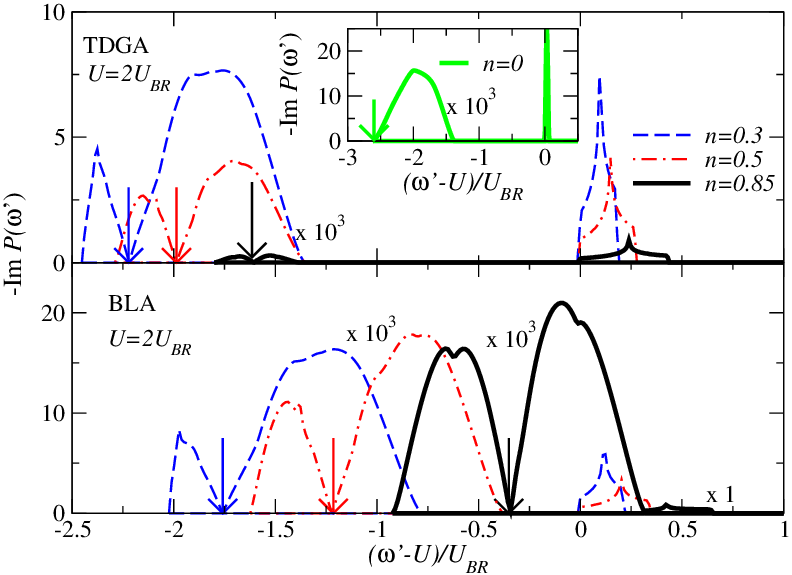}
\caption{
Local spectral function for different fillings in TDGA and BLA. Results are 
for the Hubbard model on a square lattice with nearest-neighbor hopping 
($U_{BR}=128t/\pi^2$). 
The vertical arrows indicate the position of $2\mu$, separating the addition 
part $\omega'>2\mu$ from the removal part $\omega'<2\mu$. The intensity of 
two-particle scattering states have been multiplied by $10^{3}$. 
Inset: the $n=0$ case which coincides in the two approximations.}
\label{pdw}

\vspace{-0.3cm}

\end{figure}

Fig.~\ref{figueff}(a) shows $V$ [c.f. Eq.~\eqref{eq:veff}] 
as a function of $U$ for a band of bare width $W$ and a flat 
density of states. For fillings close to 0 or 2 or small $U$, 
the effective interaction is close to the bare $U$, as expected. 
By contrast, as the Mott phase is approached, $V$ diverges. This singular 
behavior is essential for the correct description of dense systems close to 
the Mott insulator. 

For the following analysis, it is convenient to shift the 
origin of energies to eliminate the chemical potential in Eq.~\eqref{chi0}, 
i.e., we define $\omega'=\omega+2\mu$. 
Following Refs.~\cite{yan89,zha90c}, one can compute exactly 
$P({\bf Q}, \omega)$ at ${\bf Q}\equiv(\pi,..,\pi)$ for the Hubbard model with 
nearest-neighbor hopping. In this case, the full spectral function is exhausted
by a single pole at $\omega'=U$. The antibound state consist of a
doublon, i.e., an on-site pair. This provides a quick and instructive 
check of the theory. Indeed, by using that $e_{\bf k}+e_{\bf k+Q}=0$, one can 
verify that both BLA and the present theory reproduce the exact result. For
general momenta and large $U$, Eq.~\eqref{eq:prpa} has a single pole
for $\omega'\sim U$ (i.e., the antibound state) and a continuum at 
low (high) energy for $n<1$ ($n>1$). The local response is obtained as
$P_{ii}(\omega)=1/N_s \sum_{\bf q} P({\bf q},\omega)$.
Fig.~\ref{figueff}(b) shows a sketch of
${\rm Im}P_{ii}(\omega')$ with the continuum at 
$2\Sigma-z^2_0 W \le \omega' \le 2\Sigma+z^2_0 W$ and the antibound state 
at $\omega' \sim U$. The dip in the continuum at $2\mu$ separates the addition 
part for $\omega'>2\mu$ from the removal part for $\omega'<2\mu$.

The problem within the BLA is not so much the energy of the antibound
state but rather the position of the two-particle continuum, which is given 
by the HF eigenvalues. This affects the antibound state because, as
the continuum approaches the energy $U$, the antibound pair becomes less
localized in the relative coordinate and eventually disappears. 
The distance between the continuum 
($\sim 2\Sigma$) and the antibound state ($\sim U$) as a function of $U$ is
shown in Fig.~\ref{figueff}(c). The picture can be easily understood
by noticing that by rescaling the $y$ axis by 1/2 one obtains 
one-particle energies. For an almost filled or an almost empty band, as 
well as for small $U$, the position of the HF and the GA band
coincide. At half filling, the HF self-energy is $\Sigma_{HF}=U/2$ so
that $2\Sigma-U=0$. This coincides with the GA for $U<U_{BR}$, however,
for $U>U_{BR}$ the GA self-energy bifurcates in two solutions, corresponding
to infinitesimal positive and negative deviation from half filling, due
to the opening of the Mott-Hubbard gap. Thus for $U$ larger than
$U_{BR}$ and moderate filling, the HF band is close to $U$, whereas the GA 
band is well separated from it. In this case we can anticipate
quite different two-particle spectra in the two approximations. 
This dramatic difference is 
also illustrated in Fig.~\ref{figueff}(d) where the boundaries of the 
continuum with respect to the doublon energy are shown as a function of 
filling for $U=2U_{BR}$. 
The HF self-energy leads to a linear evolution. By contrast, in GA the band 
remains nearly at the same energy and narrows when $n\rightarrow 1$ due to 
correlation. At $n=1$ the band jumps due to the Mott-Hubbard gap and the 
situation reverses.
Clearly the GA continuum is always far from the antibound state whereas, 
in HF, it is generally much closer and overlaps the $\omega'=U$ line in a 
large range of filling near $n=1$. Therefore, the formation of tight antibound 
states will be much more favored in the GA case. 

Fig.~\ref{pdw} compares the local two-particle spectral function for an 
infinite two-dimensional system and $n<1$, within TDGA and BLA.
The inset shows the $n=0$ case where TDGA and BLA coincide. 
Differences occur at finite concentrations (main panel) where the line 
shapes are dominated by the antibound state at $\omega'\sim U$ (as in CS),
which is significantly stronger in the TDGA. The intensity of 
the continuum at low energies has been multiplied by $10^3$ to make the 
line shape visible. As anticipated the two-particle continuum is far from
the antibound state in GA, whereas it quickly approaches it in the BLA.

The antibound state can propagate and forms a band which gives the
width of the high-energy feature. The lower edge of this
band corresponds to ${\bf q}=(\pi,\pi)$ for $n<1$ and is at $\omega'=U$. 
For large $U$, the bandwidth is of order $t^2/U$ for $n=0$\cite{saw77} but 
becomes of order $t$ (specifically $2z_0^2|\bar e/(1-n)|$) for finite $n$,
since the kinetic energy can move a doublon at first order if there is a 
single occupied site next to it.   

The pair correlation function satisfies the sum rules:
\begin{eqnarray}
  -\frac1\pi \int_{2\mu}^{\infty} d\omega' {\rm Im}\,
  P_{ii}(\omega')&=&1-n+\langle n_{i\uparrow}n_{i\downarrow}\rangle
\label{eq:intp1}
\\
  -\frac1\pi \int_{-\infty}^{2\mu} d\omega' {\rm Im}\,P_{ii}(\omega')&=&\langle n_{i\uparrow}n_{i\downarrow}\rangle.
\label{eq:intp2}
\end{eqnarray}
This can be used to evaluate ladder corrections to the
GA or HF double occupancy. 
The fact that the area of the removal part is much larger in the BLA than in 
TDGA reflects a larger double occupancy in the former.
This is not surprising since at  zero order BLA neglects
correlations at all. Furthermore, in TDGA, as the system approaches the Mott
phase, the hopping renormalizes to zero and the system becomes more
``atomic'' like. This explains the vanishing of  two-particle
scattering states as $n\rightarrow 1$. (Clearly in an exact computation, 
a small finite double occupancy and scattering intensity will remain in the 
Mott phase). Contrary, in the BLA the system becomes more ``band'' like as 
the filling is increased due to the closing of the gap between the scattering 
states and the doublon energy. Indeed for $n=0.85$ the antibound state 
exists only for some values of the momentum. 

In order to validate our results, we have computed the exact two-particle
addition spectra for $10$ particles on a $4\times 4$ lattice with only
nearest-neighbor hopping $t$ and $U/t=15$, by using exact diagonalization.
Fig.~\ref{fig:u15} shows a comparison between the present theory and BLA. 
Here, we go back to our original variables and fix the origin of energy at 
$2\mu$. Despite the large value of the Hubbard repulsion, TDGA yields 
excellent agreement with exact diagonalization concerning the location,
width and intensity of the high-energy antibound states.
On the other hand, BLA predicts that these excitations have a much lower
energy when referenced to $2\mu$ and no clear separation with the band
states is visible (see upper-left inset). For the system under consideration,
there are three band-like two-particle energies which are very well reproduced
by TDGA in contrast with BLA.
The upper-right inset demonstrates that the double occupancy after
Eq.~\eqref{eq:intp1} is  accurate within TDGA, whereas BLA overestimates it as 
expected. 
The excellent performance of TDGA is not restricted to this particular value 
of $U$ but persists to even larger (and of course lower) on-site repulsions.

To conclude, we have presented a computation of pair fluctuations for
the Hubbard model exhibiting antibound states for large Coulomb repulsion. 
Our approximation gives reliable results even for large densities, where we 
are not aware of any accurate theory. The simplicity of the method 
suggests its application to the computation of Auger spectra on top of
realistic Gutzwiller calculations\cite{buenemann}.
Our theory can also be applied to ultra-cold fermion atoms in optical
lattices, which can be described by the Hubbard model as well\cite{hof02}. 
The possibility to observe antibound states has already been demonstrated
in the Bose case\cite{win06}.

\acknowledgments 
G.S. acknowledges financial support from the Deutsche Forschungsgemeinschaft,
F.B. and J.L. from CNR-INFM.

\end{document}